\documentclass[aps,prbbib,floatfix,floats]{revtex4}
\usepackage{graphicx,latexsym}
\usepackage{epsfig}
\usepackage{amsmath}
\usepackage{color}
\begin{document}
\title{\bf Kondo peaks and dips in the differential conductance of a multi-lead
quantum dot: Dependence on bias
conditions}
\author{M. \c{T}olea$^{1}$, I. V. Dinu$^{1}$, and  A. Aldea$^{1,2}$ }
\affiliation{\hskip-1.4cm
$^1$ National Institute of Materials Physics, POB MG-7, 77125
 Bucharest-Magurele, Romania. \\
$^2$ Institute of Theoretical Physics, Cologne University, 50937
Cologne, Germany.}

\begin{abstract}
We study the differential conductance  in the
Kondo regime of a quantum dot coupled to multiple leads. When the bias is
applied symmetrically on two of the leads ($V$ and $-V$, as usual
 in experiments), while the others are grounded, the conductance
through the biased leads always shows the expected enhancement
at {\it zero} bias. However, under asymmetrically applied bias
($V$ and $\lambda V$, with $\lambda>0$ ), a suppression - dip -
appears in the differential conductance if the asymmetry
coefficient $\lambda$ is beyond a given threshold $\lambda_0=
\sqrt[3]{1+r}$ determined by the ratio $r$ of the dot-leads
couplings. This is a  recipe to determine experimentally this
ratio which is important for the quantum-dot devices. This
finding is a direct result of the  Keldysh transport formalism.
For the illustration we use  a many-lead Anderson Hamiltonian ,
the Green functions being calculated in the Lacroix
approximation, which is generalized to the case of
nonequilibrium.
\end{abstract}
\maketitle

\section{Introduction}
The observation of the Kondo effect in quantum dots allows a
direct insight on this interesting many-body effect, whose
experimental fingerprint is the enhanced differential
conductance at $zero$ bias
which is a specific type of $zero$ bias anomaly.
The great advantage over the bulk Kondo effect is of course the
possibility to control the parameters like for instance the
coupling with the leads. Next, it was shown that even more
information can be extracted if the dot is connected to three
leads. The three lead-Kondo problem was theoretically
considered before in, e.g., \cite{LS, SunGuo,Rosch} and
experimentally realized by Leturcq et. al \cite{Leturcq,
Leturcq_pssb}. The third lead was used to read the
non-equilibrium density of states (DOS) induced by the other
two leads. Shah and Rosch \cite{Rosch} analyzed the influence of
the coupling and bias on the decoherence rate.

In this paper, we also study the many-lead quantum dots in Kondo regime,
but the focus is different : we are interested to find out  the
conditions under which the differential conductance can show a
suppression at {\it zero} bias instead of the usual enhancement.
There are well-known examples of transport properties influenced by
dips in the spectral function produced by Coulomb interactions. The
first one was the Coulomb pseudo-gap in the impurity conduction of
doped semiconductors resulting in the $T^{-1/2}$ dependence of the
hopping conductivity\cite{Efros}. Then, a zero bias anomaly
manifested as a dip in tunneling DOS was studied in tunnel
junctions\cite{Altshuler}. However, what is
conspicuous  for the three-lead quantum dots (see sketch in  Fig.2),
is that a peak-dip crossover may appear in the differential
conductance, while keeping the dot permanently in  Kondo regime. As
we show in the next section, this occurs when the bias is applied
asymmetrically on the left and right leads, and the crossover is
triggered by the degree of asymmetry.
For the sake of definiteness, let us assume that for some
voltage configuration (see sketch in  Fig.2), the left lead
receives electrons from the central lead and pumps electrons
into the right lead, the difference of the two contributions
being the total current. For the conductances, one can write
$\mathcal G_L=\mathcal G_{LC} +\mathcal G_{LR}=\mathcal G_{LC} -|\mathcal G_{LR}|$. Both
$\mathcal G_{LC}$ and $|\mathcal G_{LR}|$ must decay by
applying a voltage $V$ (as the voltage destroys the Kondo
resonance and reduces the electronic tunneling between leads), but not necessarily with the same derivative ;
then,  if $d\mathcal G_{LC}/dV > d|\mathcal G_{LR}|/dV$ the
conductance $\mathcal G_L$ will increase with $V$, giving rise
to a dip about $V=0$.

It will also be shown  that the peak-dip crossover can be used
to determine the ratio between the couplings of different leads
to the dot, and therefore it is of potential experimental use.

The main conclusions of the paper are extracted  from the current
formula given by the Keldysh formalism (Eq.1) by taking advantage of
a simple connection between the convexity of the conductance $dI/dV$
and the convexity of the density of states.

Recently, in the first two-channel Kondo experiment \cite{Potok}, a dip
generated by the competition between channels was observed in the
differential conductance, even in the case of symmetrically applied bias.
One may note that in our calculations, as in the mentioned
(theoretical and experimental) papers referring to the 3-lead problem
 \cite{LS,Leturcq, Leturcq_pssb,Rosch}, the set-up and physics are different,
and therefore the results are outside the multi-channel Kondo phenomenology.

For the illustration of the results, we shall give also
some numerical calculations using the equation of motion
solution for the Anderson Hamiltonian \cite{Lacroix},
generalized to the multi-lead and non-equilibrium case.

\section{Peak-dip crossover in differential conductance: analytical analysis}
In order to calculate the electron current through the lead $\alpha$
one can use the Keldysh formalism, and obtain \cite{LS}:
\begin{equation}
I_\alpha=-\frac{4e\Delta_\alpha}{\pi \hbar
\Delta}\int_{-\infty}^{\infty}{ImG_{dd}(\omega)
\sum_{\beta (\neq \alpha)}\Delta_\beta\big[ f(\omega-\mu_\alpha)-
f(\omega-\mu_\beta)\big]d\omega}, ~~~~~\sum_{\alpha}I_{\alpha}=0,
\end{equation}
where $\Delta_\alpha$ is the coupling of the lead $\alpha$ to the dot,
 $\Delta=\sum{\Delta_\alpha}$, and
$f(\omega-\mu_\alpha)$ is the Fermi distribution in the $\alpha$
lead.  $G_{dd}$ is the retarded Green function of the dot in the
presence of the leads. We remind that, usually, the Keldysh
formalism needs also the "lesser" Green function , but when the
leads are all connected in the same site, the lesser function can be
eliminated (except for the case of magnetic leads \cite{sw}). It was
recently proven that the same simplification occurs for a symmetric
many-site system \cite{noi}.

  In this paper we are interested in the behavior of the
conductance close to equilibrium, for the case when two of the
leads are biased and the others are grounded. In this case, all
the grounded leads have the same chemical potential (set to
$zero$) and act like one single reservoir; this reservoir will
be called "central lead" (with the chemical potential $V_C=0$).
 If we apply the potential $V_L=V$ on the left lead $"L"$ and the
potential $V_R=\lambda V$ on the right  lead $"R"$, then the
differential conductance through the lead $"L"$ can be calculated
from Eq.1  :
\begin{eqnarray}
{\mathcal G_L}(V)&=&\frac{d I_L}{d V}=\frac{4e^2\Delta_L}{\pi \hbar
\Delta}\int_{-\infty}^{\infty}{
\big[(\Delta_R+\Delta_C)ImG_{dd}(\omega)\frac{d f(\omega-eV)}{d
\omega}-\lambda\Delta_R ImG_{dd}(\omega) \frac{d f(\omega-\lambda
eV)}{d \omega}\big]d\omega}.
\end{eqnarray}
Notice that only  the
derivative of the Fermi function was considered while the
derivative of $ImG_{dd}$ versus bias was neglected, which was
proven to be a very good approximation in \cite{LS}.
 At low temperatures, when the derivatives of the Fermi
functions approximate delta functions,  the
above formula turns into :
\begin{equation}
{\mathcal G_L}(V)=-\frac{4e^2\Delta_L}{\pi \hbar \Delta}
[(\Delta_R+\Delta_C)ImG_{dd}(eV)-\lambda\Delta_R ImG_{dd}(\lambda eV)]
\end{equation}
and similarly:
\begin{eqnarray}
{\mathcal G_R}(V)&=&\frac{d I_R}{d V}=-\frac{4e^2\Delta_R}{\pi
\hbar \Delta}
[\lambda(\Delta_L+\Delta_C)ImG_{dd}(\lambda eV)-\Delta_L ImG_{dd}(eV)]\\
{\mathcal G_C}(V)&=&\frac{d I_C}{d V}=-\frac{4e^2\Delta_C}{\pi
\hbar \Delta}
[-\Delta_L ImG_{dd}(eV)-\lambda \Delta_R ImG_{dd}(\lambda eV)]
\end{eqnarray}
 The conductance $\mathcal G_{L,R,C}$, defined in this way
is a quantity that can be measured experimentally and helps for determining
the coupling ratios as we shall see below.

We suggest now a way to search for Kondo dips in the
differential conductance. One has to compute the second
derivative of $\mathcal G_{L,R,C}$ for $V=0$ which contains the
convexity information (in case the first derivative vanishes,
the convexity determines if one has a peak or a dip). The
second derivative of the differential conductance can be
expressed as:
\begin{eqnarray}
{\mathcal G_L}''(0)=-\frac{4e^4\Delta_L\Delta_R}{\pi \hbar \Delta}
ImG_{dd}''(0)\big[(1+\frac{\Delta_C}{\Delta_R})-\lambda^3\big]\\
{\mathcal G_R}''(0)=-\frac{4e^4\Delta_L\Delta_R}{\pi \hbar \Delta}
ImG_{dd}''(0)\big[(1+\frac{\Delta_C}{\Delta_R})\lambda^3-1\big]\\
{\mathcal G_C}''(0)=\frac{4e^4\Delta_C\Delta_R}{\pi \hbar \Delta}
ImG_{dd}''(0)\big[\frac{\Delta_L}{\Delta_R}+\lambda^3\big];
\end{eqnarray}
 In the above formulas $-ImG_{dd}''(0)$ is the second derivative of
the spectral function at equilibrium (i.e. for $V=0$) and is always
negative for $T<T_K$ indicating the presence of the Kondo peak in
the density of states.  This
property will be used to determine the peak or dip behavior of the
conductance.

The distribution (and sign) of the conductances in the
three-lead device is controlled by the coupling to the ground
lead $\Delta_C$. In the mentioned experiments
\cite{Leturcq,Leturcq_pssb} , when the aim was to read the
non-equilibrium DoS, $\Delta_C$ is very small compared to the
other couplings; in this case the currents L and R have always
opposite signs, the same being true for the second derivatives
(6-7); the conclusion is that both conductances are maximum in
module at zero bias. However, for large coupling $\Delta_C$  the
above formulae indicate the possibility of a more complex
behavior of the conductance.

We identify first the case
$\lambda=-1$ , as being the one commonly used, with the gate
potentials symmetrically applied.
 In Eq.6, the straight parenthesis on the r.h.s is always positive for
negative $\lambda$, so one may conclude that ${\mathcal G_L}''(0)$
has the same sign as $-ImG_{dd}''(0)$, which is negative, indicating
a peak also in the differential conductance ${\mathcal G_L}$.
For negative $\lambda$, it turns out from Eqs.3,4 that $\mathcal G_L>0$ and
$\mathcal G_R<0$,  meaning that the left and right currents flow in
opposite directions.
Then the above argument holds  for the module of ${\mathcal G_R}$
which is also maximum at zero bias.

Interestingly, for $\Delta_L=\Delta_R$ and $\lambda=-1$,  ${\mathcal
G_C}$ is not zero as a consequence of the spectral asymmetry
$ImG_{dd}(eV)\ne ImG_{dd}(-eV)$ (see Eq.5, and also notice the different
heights of the side peaks in Fig.1). The same non-zero value of the
current through the grounded lead was discussed in \cite{Flensberg},
for the sequential tunneling case, and attributed to the
electron-hole asymmetry. Under the above mentioned conditions ${\mathcal
G_C}''(V=0)=0$ indicating a saddle point. In this case, the conductance
through the central lead does not show a peak shape,
being instead antisymmetrical about $V=0$ .
\begin{figure}[ht]
\centering
\epsfxsize=0.5\textwidth \epsfbox{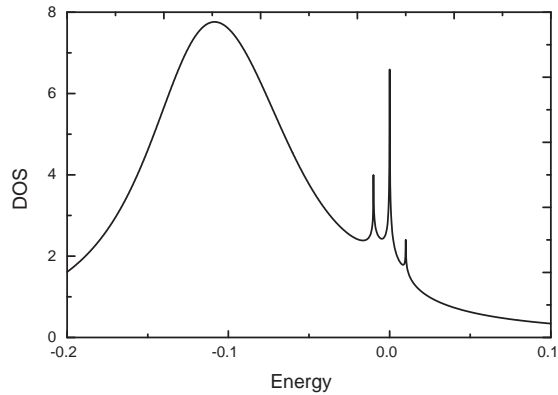}
\caption { The dot density of states  for the case of symmetrically applied bias
($\lambda=-1$). The three Kondo peaks correspond to the chemical
potentials in the three leads. Notice that the side peaks are different in height.
The parameters are:$V=0.01$, $\epsilon_d=-0.12$,$\Delta_R=\Delta_L=0.01$, $\Delta_C=0.02$
 $\Delta_C/\Delta_L=2$, $\Delta_R=\Delta_L=0.01$. The temperature is $T=T_K/10$, and $T_K=3.2\cdot10^{-4}$.}
\end{figure}

\begin{figure}[ht]
\centering
\epsfxsize=0.5\textwidth \epsfbox{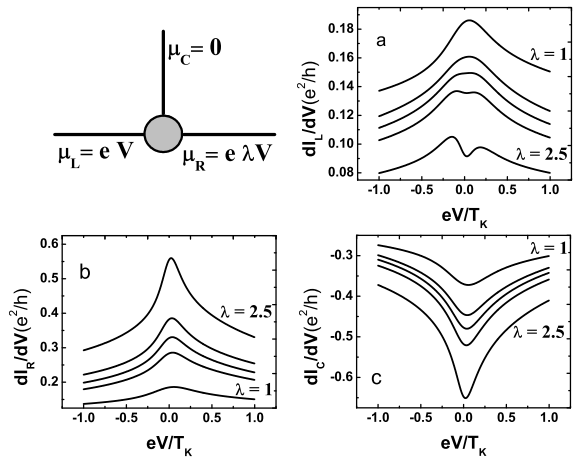}
\caption {Sketch of a quantum dot connected to three leads, with the bias
asymmetrically applied on the left and right leads
($V$ and $\lambda V$, with $\lambda = 1/1.4/1.58/1.8/2.5$).
(a) The differential conductance through the left lead changes from peak to dip
(around {\it zero} bias). The value of the asymmetry parameter $\lambda$
where the crossover happens is $\lambda_0=\sqrt[3]{1+\Delta_C/\Delta_L}=1.58$.
(b,c) Differential conductance through the right and central lead show a Kondo
peak behavior being maximum in module at $V=0$
The other parameters : $\Delta_R=\Delta_L=0.075$,$\Delta_C/\Delta_L=3$,
 $\epsilon_d=-0.12$, $T=T_K/10$, and $T_K=3.2\cdot10^{-4}$.}
\end{figure}
A different situation occurs in the case $\lambda>0$. It is
obvious  that, without loss of generality, we may consider actually
 $\lambda\geq 1$  meaning that, by convention, the smaller
bias is applied on the left lead. One may now conclude from Eq.6
that the peak/dip behavior in the left lead is completely
determined by the sign of $(1+\Delta_C/\Delta_R)-\lambda^3$. This
result may be of potential interest for experimentalists. In
particular, if one finds the crossover value $\lambda_0=
\sqrt[3]{1+\Delta_C/\Delta_R}$ (when the peak changes into dip), the
ratio $\Delta_C/\Delta_R$ can be immediately extracted. The
mentioned coupling ratio is otherwise inaccessible to direct
measurement. Such a peak-dip crossover can be seen in the Fig.2a.
The numerical plot (see below calculation details)
 indicates that the dip is considerably narrower than
the Kondo temperature, and is situated in the middle of a larger
peak. This means that even if the conductance through the left lead
initially increases, it will eventually start to decrease for a larger
bias. It is an expected behavior, since for large biases all Kondo
correlations are killed and all the conductances approach the (much
lower) non-Kondo limit.

The conductances through
the right and central leads have the usual peak behavior (see Fig.2b,c).
One has to pay attention to the fact that the conductance through the central lead
is negative due to the sign convention regarding the
currents direction, but it is maximum in module at $zero$ bias, and should be
considered as a Kondo peak.

We stress that the formulas Eqs.6-8 are quite general, and also hold
outside the Kondo regime (for high temperatures or weak couplings),
but in this case they are not very insightful. The variation of the
differential conductance is very smooth  outside the
Kondo regime, and it is difficult to determine its convexity
unambiguously.
\section{Application: numerical results for the multi-lead Anderson model}
In order to illustrate the results, we shall calculate the
differential conductance for the   multi-lead Anderson model:
\begin{equation}
H=\sum_\sigma{\epsilon_d d_\sigma^\dagger d_\sigma}
+Ud^\dagger_\uparrow d_\uparrow d^\dagger_\downarrow d_\downarrow
+\sum_{\alpha,k,\sigma}\epsilon_{k \sigma}{c^\dagger_{k \alpha  \sigma}c_{k \alpha \sigma}}
+\sum_{\alpha,k,\sigma}V_\alpha (c^\dagger_{k \alpha  \sigma}d_\sigma+H.c.),
\end{equation}
where $\epsilon_d$ is the energy of the dot level, $U$ is the
Coulomb repulsion and, for instance, $c_{k \alpha  \sigma}$
annihilates an electron with momentum $k$ and spin $\sigma$ in the
$\alpha$ lead. The last term describes the coupling between the dot
and  leads. The quantity needed in the current formula Eq.1 is the
dot retarded Green function
$G_{dd,\sigma}=<<d_{\sigma},d_{\sigma}^\dagger>>$, which will be
calculated in the Lacroix \cite{Lacroix} approach. We write the
equation of motion for the
 Green function and perform a decoupling approximation at the
third step in order to close the system of equations. In our case,
we consider three leads (reservoirs) instead of one as in
\cite{Lacroix}, the generalization being straightforward.
The resulting formula, in the $U\rightarrow \infty$ limit, reads:
\begin{equation}
G_{dd,\sigma}(\omega+i0)=\frac{1-\frac{<n_d>}{2}-\sum_{k,\alpha}
V_\alpha\frac{<d^\dagger_{-\sigma}c_{k\alpha
-\sigma}>}{\omega-\epsilon_k}}
{\omega-E_0-\sum_{k,\alpha}\frac{V_\alpha^2}{\omega-\epsilon_k}-
\sum_{k,q,\alpha,\beta}V_\alpha
V_\beta\frac{<c^\dagger_{q\alpha-\sigma}c_{k\beta-\sigma}>}{\omega-\epsilon_k}
+\sum_{k,\alpha}\frac{V_\alpha^2}{\omega-\epsilon_k}
\sum_{k,\alpha}V_\alpha\frac{<d^\dagger_{-\sigma}c_{k\alpha-\sigma}>}
{\omega-\epsilon_k}}
\end{equation}

Since we work in a non-magnetic case, the subscript $\sigma$ can be
omitted in the Green function notation. It comes now into question
how to treat the averages of the type
$<c^\dagger_{q\alpha-\sigma}c_{k\beta-\sigma}>$, for instance, where
the two operators refer to electrons in different leads. For
equilibrium, when all the Fermi levels are equal, one can use the
fluctuation-dissipation theorem which relates the average to the
corresponding Green function:
\begin{equation}
<c^\dagger_{q\alpha-\sigma}c_{k\beta-\sigma}>=
-\frac{1}{\pi}\int{f(\omega)Im<<c_{k\beta-\sigma}c^\dagger_{q\alpha-\sigma}>>
d\omega},
\end{equation}
In non-equilibrium, the Fermi functions $f_{\alpha}$ are
different in the three leads $\alpha=L,R,C$, and for the
calculation of the mentioned average we propose
\begin{equation}
<c^\dagger_{q\alpha-\sigma}c_{k\beta-\sigma}>=-(1/2\pi)
\int{(f_\alpha(\omega)+f_\beta(\omega))
Im<<c_{k\beta-\sigma}c^\dagger_{q\alpha-\sigma}>>d\omega}.
\end{equation}
In the calculations there also appear averages of the type
$<d^\dagger_{-\sigma}c_{k\alpha -\sigma}>$, so that a distribution
function on the dot is also needed. For the dot we shall use the
weighted mean  $f_d=\sum_{\alpha} \Delta_\alpha f_\alpha/\Delta$.
This intuitive expression of $f_d$ has been used in, e.g. \cite{Haug, bulka}, and we
have checked that the present results, i.e. existence of the dip and its
temperature dependance, are stable with respect to different ways
to introduce the non-equilibrium.

These non-equilibrium aspects being settled, we continue along the
Lacroix line, and with the definitions
$\Delta=\sum\Delta_\alpha=i\sum_{k,\alpha}\frac{V_\alpha^2}
{\omega-\epsilon_k}=\frac{i\pi}{2D}\sum_{\alpha}V_\alpha^2$,
$A_\alpha=-\frac{\Delta_\alpha}{\pi}\int{\frac{G^*_d(\omega')
f_\alpha(\omega')}{\omega'-\omega-i0}d\omega'}$ , and
$B_\alpha=\frac{\Delta_\alpha}{\pi}\int{\frac{f_\alpha(\omega')}
{\omega'-\omega-i0}d\omega'}$ (where $D$ is the half-bandwidth and
is taken as energy unit)
the following expression is obtained :
\begin{equation}
G_{dd}(\omega+i0)=\frac{1-\frac{<n_d>}{2}+G_{dd}^*(\omega+i0)
\sum_\alpha{B_\alpha}}{\omega-E_0+i\Delta+\sum_\alpha{B_\alpha}
+G_{dd}^*(\omega+i0)2i\Delta\sum_{\alpha}{B_\alpha}}.
\end{equation}

In getting this expression we use the approximation
$A_{\alpha}\approx -G_{dd}^*B_{\alpha}$ , which was  proposed first
by Lacroix for the Kondo regime, and later used also in
\cite{Meir05,noi}. Eq.13 can be easily solved. First, one conjugates
the equation, and after that $G_d^*(\omega+i0)$ is replaced in the
initial formula to obtain a simple algebraic second degree equation
for $G_d(\omega+i0)$, only one of the solutions being physical. The
solution is analytic, but is not included here because its
lengthy. It is important to mention that this solution of the EOM
allows to address the low temperature regime ($T\ll T_K$), which is
not possible with the more simple decoupling scheme in \cite{SunGuo}.

\begin{figure}[ht]
\centering
\epsfxsize=0.5\textwidth \epsfbox{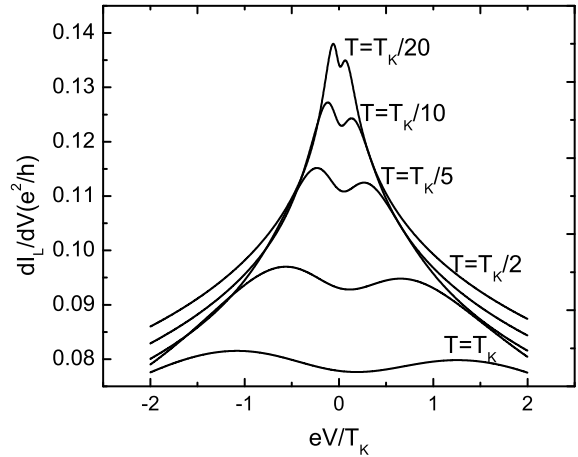}
\caption {Evolution of the Kondo dip with temperature; $\lambda=2$; the other
parameters as in Fig.2}
\end{figure}

Fig.3 shows that the Kondo-type temperature
 dependance of the conductance (i.e., decrease of  with increasing temperature)
occurs in the range $eV/T_K \in [-0.5,0.5]$ ($T_k$ being calculated as
in \cite{haldane}). This range is in fact dependent on $\lambda$
and we  checked that it becomes narrower when $\lambda$ increases.
\begin{figure}[ht]
\centering \epsfxsize=0.5\textwidth \epsfbox{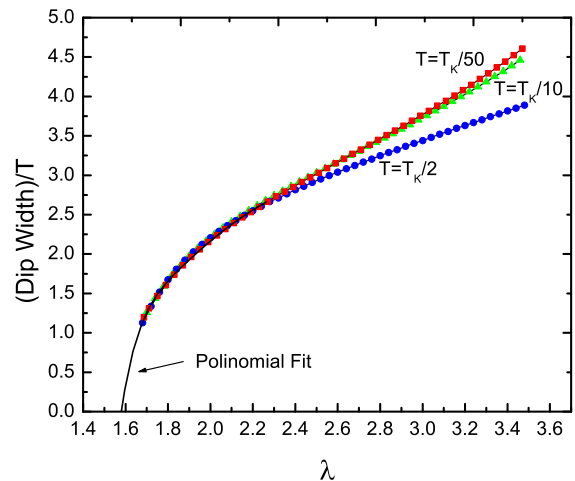} \caption{The
dip width scaled with temperature, as function of the asymmetry
parameter $\lambda$, for three different temperatures below $T_K$.
The linear temperature dependence is obvious for $\lambda\lesssim
2.4$, where the three scaled curves coincide. The dip vanishes at $\lambda_0=1.58$ which can be obtained  by
extrapolation.}
\end{figure}

One can also notice that the width of the dip increases with the temperature.
Actually the temperature dependance of the conductance comes from the
spectral function $Im G_{dd}(V=0,T)$.
Let us define the width of the dip $w (\lambda,T)$ as the distance
between the two peaks. By scaling the width  with temperature the
data from Fig.3 generates Fig.4 which shows that, for not too large
values of the parameter $\lambda$,  the width behaves like:

\begin{equation}
w(\lambda,T)= f (\lambda)~T, ~~~with~~~ f(\lambda=\lambda_0)=0.
\end{equation}
For $\lambda$ close to $\lambda_0$ the error in the calculation of
the dip width becomes large, and we suggest a numerical fit to
obtain the dip-peak crossover point $\lambda_0$. The polynomial fit
gives a value of $\lambda_0$ very close to the exact result
$\lambda_0=1.58$.

\section{Conclusions}
In conclusion, we have studied the non-equilibrium transport through
a quantum dot connected to three reservoirs. In Kondo regime, if
the potentials are applied asymmetrically on the leads (i.e. $V_R=
\lambda V_L, V_C=0, \lambda\geq1$) a peak-dip crossover of the
differential conductance occurs in left lead at zero bias. The proof
is based on the connection between the convexity of the differential
conductance and the convexity of the density of states at $V=0$ and
is independent of the Hamiltonian model. The ratios of the lead-dot
couplings ($\Delta_L/\Delta_C,\Delta_R/\Delta_C$) can be obtained if
the crossover value $\lambda_0 $  is determined experimentally.
\acknowledgements
We acknowledge support from CEEX-Research Programme and Sonderforschungsbereich
608 at the  Institute of Theoretical Physics, University of Cologne.

\end{document}